\documentstyle{article}

\title{\bf Do we live in an anthropic universe? }

\author{Domingos Soares \\ \\ {Physics Department}\\
{Federal University of Minas Gerais} \\   
{Belo Horizonte, Brazil} } 

\date {September 26, 2002}

\begin{document}

\maketitle

\begin{abstract}
I cast doubt upon the desired consistency between the anthropic principle 
and modern cosmology. 
\end{abstract}

Amongst all possible universes we live in one that {\it deserves} us. 
This is what could be called the {\it naive version} of the anthropic 
principle. At what extent this view is consistent with modern scientific 
results obtained from theoretical and observational work in cosmology? 
 
The anthropic principle was originally put forward by the cosmologist Brandon 
Carter (1974) with the statement that `our location in 
the universe is necessarily privileged to the extent of being compatible 
with our existence as {\it observers}'. (The italic is mine). 
The definite status as a consensual principle of nature has 
been crowned with the thorough account of its implications in seemingly 
unpaired areas of human knowledge such as philosophy, quantum mechanics, 
cosmology, biochemistry, the search for extraterrestrial life and ultimately 
the future of the universe, by John D. Barrow and Frank J. Tipler, in the now 
classical book entitled {\it The Anthropic Cosmological Principle} (Barrow 
\& Tipler 1986 but see Soares 2001).

Since then, two major achievements in cosmology lie on our pathway, two 
brilliant milestones. On the theoretical side, Alan Guth invented the 
inflationary theory (Guth 1997) in the early 80's, and on the observational 
side, the first results from the Cosmic Background Explorer 
satellite were published in the early 90's 
\footnote{See COBE's homepage at http://space.gsfc.nasa.gov/astro/cobe/}.
The expanding 
universe paradigm gained strength with renewed blood from these sources. 
Two recent reviews by Michael S. Turner (2002) and Max Tegmark (2002)  
give a clear picture of the present situation. The evidence for a flat 
global topology comes from both inflation and measurements of the anisotropy 
of the cosmic microwave background on angular scales of about 1 degree. The 
measurements were triggered by COBE's spectacular results, from a plethora of 
satellite and balloon experiments (see Tegmark 2002). 

Current cosmological models  should be at least reassuring of an anthropic 
universe. But, what does the cosmic budget tell us? Following 
Turner one has:
\begin{itemize}
\item Bright stars: 0.5\%
\item Baryonic dark matter: 3.5\% 
\item Nonbaryonic dark matter: 30\%
\item Dark energy: 66\%
\end{itemize}
Flatness requires that everything adds up to 100\% of the closure density. 
Except for a half per cent of visible, ordinary, {\it observable} matter, 
we are left with dark, exquisite, {\it unobservable} stuff. 

Now, back right to the beginning: 
aren't we in the {\it wrong} universe?

\section*{References}
\begin{description}
\item{Barrow, J.D. \& Tipler, F.J.} 1986, The Anthropic Cosmological 
Principle (Oxford University Press, Oxford)
\item{Carter, B.} 1974,  in  Confrontation of Cosmological Theories with 
Observation, ed. M.S. Longair (Reidel, Dordrecht), p. 291
\item{Guth, A.H.} 1997, The Inflationary Universe: The Quest for a New 
Theory of Cosmic Origins (Addison-Wesley Publishing Company, Reading) 
\item{Soares, D.S.L.} 2001,  arXiv:astro-ph/0108180
\item{Tegmark, M.} 2002,  arXiv:astro-ph/0207199
\item{Turner, M.S.} 2002, arXiv:astro-ph/0207297

\end{description}

\end{document}